\documentclass[%
 reprint,
showpacs,
 amsmath,amssymb,
 aps,prl
]{revtex4-1}

\usepackage{graphicx}
\usepackage{epsfig}
\usepackage{dcolumn}
\usepackage{bm}
\usepackage{newfloat}
\DeclareFloatingEnvironment[name={S}]{suppfig}

\begin{document}


\title{Nondestructive imaging of an ultracold lattice gas}
\author{Y. S. Patil, L. M. Aycock, S. Chakram and M. Vengalattore}
 \affiliation{Laboratory of Atomic and Solid State Physics, Cornell University, Ithaca, NY 14853}
\email{mukundv@cornell.edu}


\begin{abstract}
We demonstrate the nondestructive imaging of a lattice gas of ultracold bosons. Atomic fluorescence is induced in the simultaneous presence of degenerate Raman sideband cooling. The combined influence of these processes controllably cycles an atom between a dark state and a fluorescing state while eliminating heating and loss. Through spatially resolved sideband spectroscopy following the imaging sequence, we demonstrate the efficacy of this imaging technique in various regimes of lattice depth and fluorescence acquisition rate. Our work provides an important extension of quantum gas imaging to the nondestructive detection, control and manipulation of atoms in optical lattices. In addition, our technique can also be extended to atomic species that are less amenable to molasses-based lattice imaging. 
\end{abstract}

\pacs{37.10.Jk,67.85.Hj,03.75.Lm,03.75.-b,03.67.-a}

\maketitle

The creation, control and manipulation of ultracold atomic gases in tailored optical potentials has spurred enormous interest in harnessing these mesoscopic quantum systems to the realization of ultracold analogues of correlated electronic materials \cite{lewenstein2007, blochrmp2008}, studies of non-equilibrium dynamics in isolated quantum many-body systems \cite{polkovnikov2011} and quantum metrology \cite{bloom2014}. The dilute nature of these gases and the weak interactions impose stringent restrictions on the energy, entropy and means of manipulating and probing these systems. This has led to the development of novel techniques to cool and image these gases at ever increasing levels of precision and resolution. In this context, the {\em in situ} imaging of lattice gases at high spatial resolution has emerged as a powerful tool for the study of Hubbard models \cite{bakr2009,gemelke2009, wurtz2009, sherson2010} and quantum information processing \cite{nelson2007}. 

Here, we demonstrate a nondestructive imaging technique for ultracold lattice gases. The scheme relies on extracting fluorescence from the atoms while simultaneously cooling them to the lowest band of the lattice via Raman sideband cooling (RSC)\cite{vuletic1998,hamann1998,han2000}. Through a combination of sideband spectroscopy and time-of-flight measurements, we demonstrate broad regimes of fluorescence acquisition rates and lattice depths for which the imaging scheme preserves the spatial location, the spin state and the vibrational occupancy of the lattice gas. 

The principle of the imaging sequence is depicted in Fig. 1. Raman sideband cooling is employed to cool individual atoms within an optical lattice to the lowest vibrational band while simultaneously pumping the atoms to the high field seeking state. In the case of $^{87}$Rb atoms used in our study, this state is denoted by $|g \rangle = |F=1, m_F=1; \nu=0 \rangle$ where $\nu$ is the vibrational state of the atom within a lattice site. Importantly, this state is a dark state with respect to the optical fields used for Raman cooling. As such, the atoms do not emit any fluorescence while in this ground state. Fluorescence is induced in these atoms by shining a circularly polarized ($\sigma_-$) beam resonant with the $F=1 \rightarrow F'=0$ (D2) transition. Simultaneous use of RSC mitigates the increase in temperature caused by this fluorescence beam by cycling the atoms back to $|g \rangle$. Due to this cycling, fluorescence can be repeatedly extracted from the atomic distribution while leaving the atom in its original state. 

For the studies described below, we use 3D optical lattices that are typically detuned $2 \pi \times 160$ GHz from the $F=1 \rightarrow F'$ (D2) transition of $^{87}$Rb. The lattice provides both the confining potential as well as the coherent two-photon coupling required for sideband cooling \cite{kerman2000}. In the absence of RSC, we measure a heating rate of 11 nK/ms due to the photon scattering from the near-resonant lattice. While this does not pose a limitation for the studies described in this work, this heating can be significantly reduced by employing separate optical fields to provide the lattice confinement and the Raman coupling. 

Atoms are loaded into this lattice and initialized in the ground state $|g \rangle$ by a 10 ms period of RSC. Based on measurements of the atomic density within the lattice, we estimate filling fractions on the order of $f = $0.20 - 0.25. The average vibrational occupation number is measured using sideband spectroscopy to be $\langle n \rangle \leq 0.01$ for the entire range of lattice depths studied here. Fluorescence images are acquired by switching on the fluorescence beam at a variable intensity. The images are acquired within exposure times of up to 30 ms following which the number of atoms and temperature of the atomic distribution are measured using a combination of time-of-flight absorption imaging and sideband spectroscopy. 

We perform sideband spectroscopy to accurately quantify local changes in temperature due to fluorescence imaging. For this, we employ a pair of counter-propagating beams detuned $2 \pi \times 7.5$ GHz from the $F=1 \rightarrow F'$ (D2) transition of $^{87}$Rb. The beams are focused to an approximate waist of 8 $\mu$m. The measured sideband asymmetry \cite{monroe1995} allows a local extraction  of the vibrational occupation number (Fig. 2). An oblique orientation of these beams with respect to the lattice coordinates ensures sensitivity of the sideband spectra to atomic motion in all three dimensions. The two-photon pulses are typically $500 \, \mu$s in duration with typical pump (probe) powers of 10 $\mu$W (20 nW). We have verified that the vibrational occupation number extracted from the sideband spectra is consistent with temperatures measured by time-of-flight imaging following a rapid ($< 1 \, \mu$s) extinction of the lattice \cite{toffootnote}. Also, the observed width of the sidebands is consistent with our estimate of the coherent Raman coupling induced by the near-resonant lattice. 

\begin{figure}[t]
\includegraphics[width=3.50in]{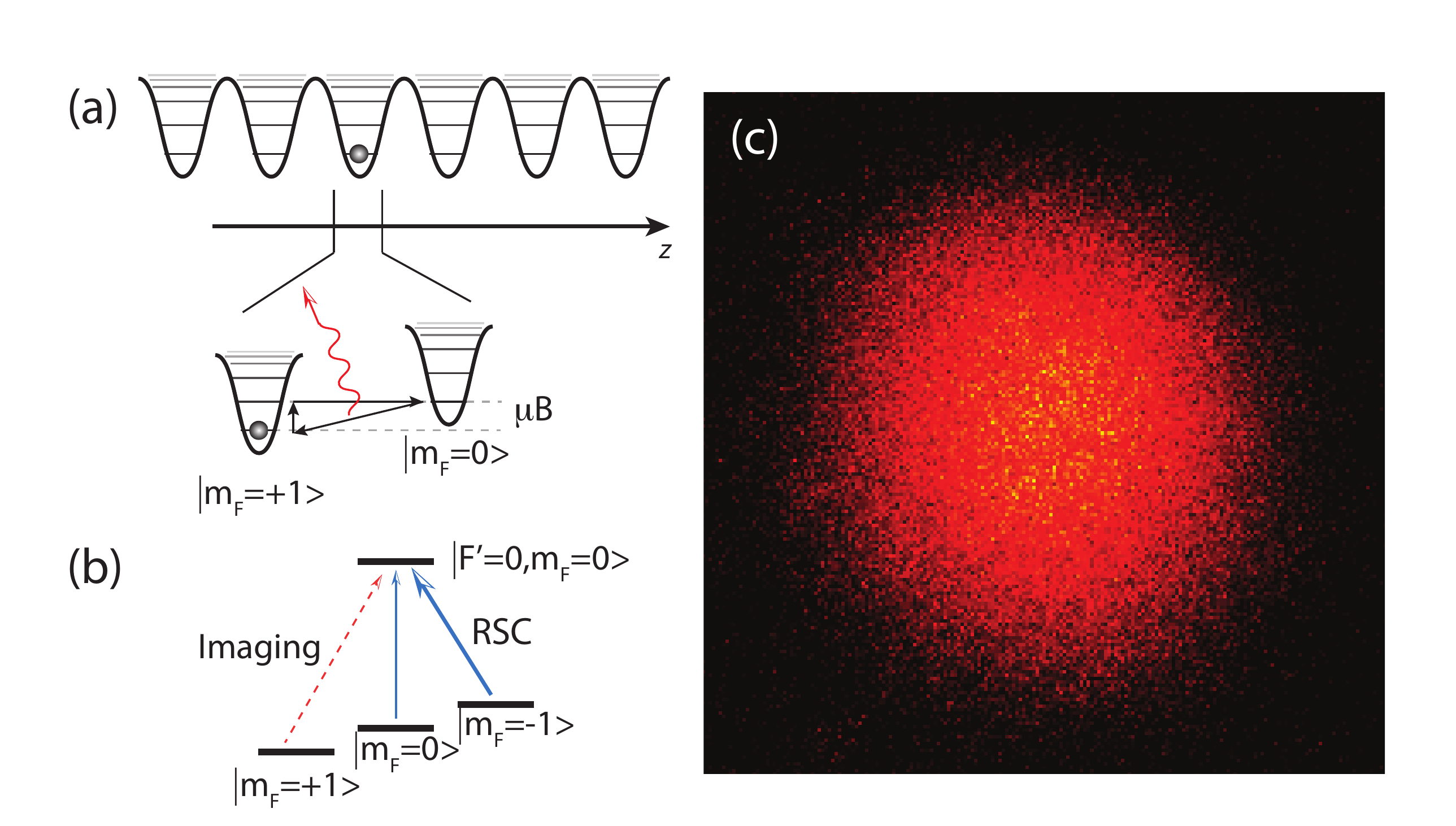}
\caption{(a) Lattice imaging scheme : An atom within a lattice site is cooled to the ground state $|g\rangle \equiv |F=1,m_F=+1; \nu = 0 \rangle$ via RSC. An auxiliary `imaging' beam promotes the atom out of this state to a fluorescing state which is subsequently cooled back to $|g\rangle$. Repeated cycles of this process extract fluorescence from the atom while continually restoring the atom to $|g \rangle$ ; (b) The near-resonance optical fields used in the imaging sequence. A cooling beam (RSC) with $\sigma_+$ and $\pi$ components cools and optically pumps the atom into the dark state $|g \rangle$. A $\sigma_-$ beam induces fluorescence by bringing the atom out of the dark state. (c) Raman fluorescence image of a gas of $1.5 \times 10^6$ atoms obtained within 15 ms. The field of view is $250 \mu$m $\times$ $250 \mu$m.}
\label{Fig:Fig2}
\end{figure}

Our imaging scheme, as constructed, relies on the competition between two processes : Atomic fluorescence at a rate $\Gamma_f$ that yields spatial information about the atomic distribution, and RSC at a cooling rate $\Gamma_{RSC}$ that serves to cool the atoms back to the ground state within each lattice site. While the former depends solely on the intensity of the fluorescence beam, the latter is given by $\Gamma_{RSC} \sim \Gamma_{opt} \times \Omega_R^2/(\Gamma^2_{opt}  + 2 \Omega_R^2)$ where $\Omega_R$ is the coherent Raman coupling between the states $|m_F, \nu \rangle$ and $|m_F-1, \nu - 1 \rangle$, and $\Gamma_{opt}$ is the rate of optical pumping to the $|m_F = +1 \rangle$ state \cite{cct1998}. As the fluorescence rate is increased significantly beyond the cooling rate, atoms can be promoted to higher bands within the lattice and can tunnel to neighboring sites. In addition to modifying the atomic distribution, such tunneling can also lead to multiply occupied lattice sites and subsequent loss due to light-induced collisions. 

To identify the regimes of imaging where the scheme is nondestructive, we use light-induced collisional loss as a diagnostic tool to monitor atomic tunneling across lattice sites. Further, in order to clearly demarcate atomic dynamics due to the fluorescence pulse from that due to RSC, we employ a pulsed imaging sequence wherein the fluorescence pulse and RSC are employed in rapid succession with a variable duty cycle. As expected, the average vibrational occupation number measured at the end of the fluorescence pulse grows with increasing fluorescence rate (Fig. 3(a)). However, RSC is very efficient at cooling the atoms back to the ground state at the end of each cycle. At the end of each RSC cycle, we measure average vibrational occupation numbers ($\sim 0.01$) that are, within our measurement uncertainty, indistinguishable from those measured in the absence of the fluorescence pulse (Fig. 2). The typical measured RSC cooling rates of 13 $\mu$K/ms are also consistent with that estimated based on the intensities of the lattice and optical pumping beams. 

\begin{figure}[t]
\includegraphics[width=3.10in]{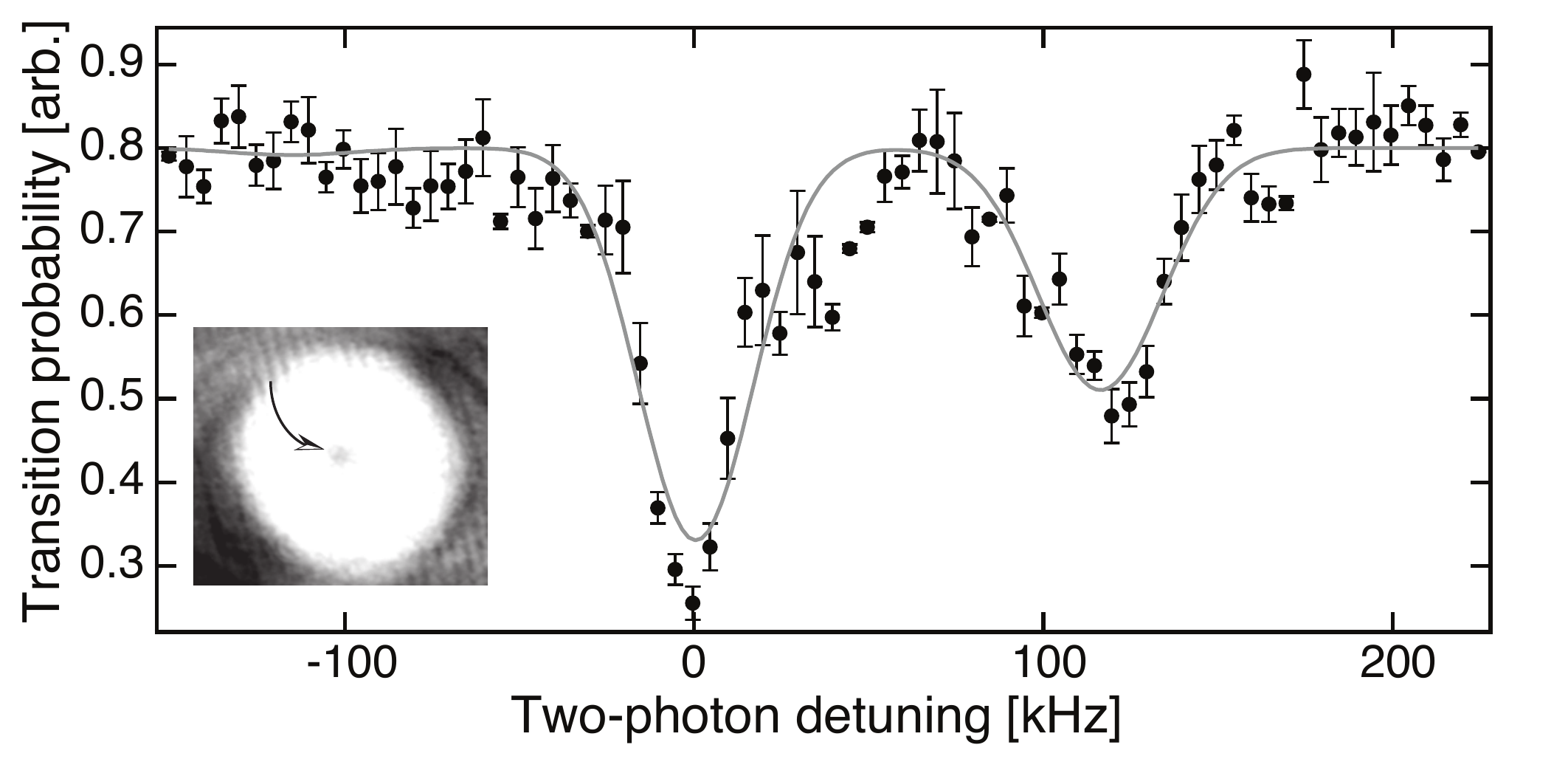}
\caption{Spatially resolved sideband spectroscopy of the lattice gas following the imaging sequence yielding $\langle n \rangle = 0.01^{+0.03}_{-0.01}$. Inset: A time-of-flight absorption image of the ultracold gas following an intense interrogation pulse at the two-photon resonance. The divot near the center of the atomic distribution shows the location and relative size of the beams used for sideband spectroscopy. The field of view is 600 $\mu$m $\times$ 600 $\mu$m.}
\label{Fig:Fig2}
\end{figure}

While simultaneous cooling during fluorescence acquisition leaves the final vibrational occupation unaltered, the transient increase in temperature during the fluorescence pulse can cause tunneling of atoms to neighbouring lattice sites. This tunneling rate depends sensitively on both the average vibrational occupation number as well as the lattice depth $U_0$, typically parametrized in units of the recoil energy $E_r = \hbar^2 k^2/2 m$. At low rates of fluorescence acquisition \cite{ratenote}, we observe that the total number of atoms is left unchanged subsequent to the imaging sequence indicating a negligible level of tunneling across sites. Beyond a certain fluorescence rate $\Gamma_{f,max}$, we observe two-body loss indicating the onset of tunneling of atoms  (Fig. 3 (b)).  As indicated by the rapid decrease of atoms for fluorescence rates past this maximal value, two-body loss is a very sensitive measure of the tunneling rates induced by the imaging sequence (see also \cite{syassen2008, yan2013}). The temporal evolution of the atom number following a brief, intense fluorescence pulse indicates that RSC cools and binds the atoms to the ground state of a lattice site within 100 $\mu$s (see inset of Fig. 3(b)), again consistent with measurements of the cooling rate. 

We have performed Monte-Carlo simulations of the imaging process that accurately capture the sensitivity of two-body loss to tunneling events and the threshold behavior arising from the competition of imaging, RSC and tunneling. For the filling fractions used in this work (0.20 - 0.25), the measured critical fluorescence rate, $\Gamma_{f,max}$, as identified by the onset of light-induced loss, is within 20\% of the critical fluorescence rate for the onset of tunneling.
We further find that the filling fraction needs to be reduced by more than an order of magnitude before there is a significant probability of tunneling events that do not lead to measurable loss. These findings justify the correspondence between the onset of tunneling in the lattice gas and our measured onset of two-body loss. 

\begin{figure}[t]
\includegraphics[width=2.9in]{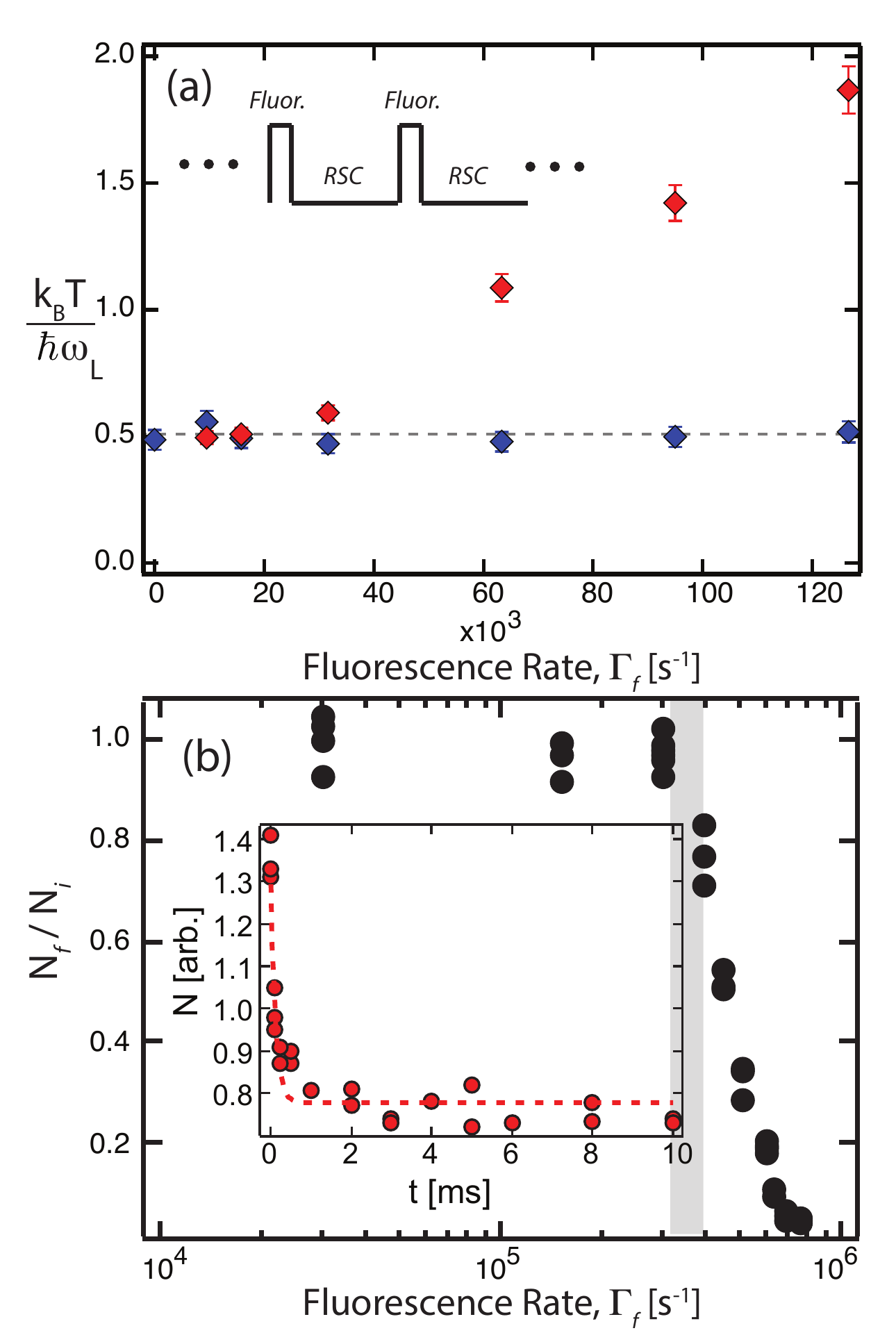}
\caption{Regimes of fluorescence acquisition rates for non-destructive imaging.  (a) Measured temperature of the lattice gas in a pulsed imaging sequence. The temperature during the fluorescence pulse grows (red) with increasing fluorescence rates while RSC rapidly cools the atoms back to the ground state (blue). (b) Measured atom number following the imaging sequence. At low fluorescence rates, the atom number is conserved indicating negligible levels of tunneling across sites. As the fluorescence rate is increased, the increasing temperature during the fluorescence pulse causes tunneling followed by light-induced loss. The shaded area represents the critical fluorescence rate for the onset of tunneling as identified by our measurements of light-induced loss. Inset: Evolution of atom number immediately following an intense fluorescence pulse ($\Gamma_f = 6 \times 10^5$ s$^{-1}$) shows that RSC quickly (within 100 $\mu$s) binds the atoms to the ground state of a lattice site thereby drastically suppressing tunneling. }
\label{Fig:Fig3}
\end{figure}

Similar considerations apply to the imaging of atoms in shallow optical lattices (Fig. 4). In this case, the rates of tunneling grow exponentially with decreasing lattice depth \cite{jaksch1998, zwerger2003}. This leads to a reduction of the maximal fluorescence rates that can be used while constraining atomic motion. As expected, an estimate of this maximum allowable fluorescence rate shows an exponential decrease with lower lattice depths (see inset of Fig. 4). Importantly, we see that fluorescence acquistion rates greater than $10^4$ photons/s per atom are possible even for lattice depths around 15 recoil energies. This makes possible the use of this imaging technique to study lattice gases in regimes where coherent tunneling of atoms within the lowest band occurs on experimentally relevant timescales. In addition, it augurs the intriguing possibility of using this imaging scheme to influence or exert spatial control over such coherent tunneling processes. Nondestructive imaging of atoms in even lower lattice depths could be made possible by increasing the Raman cooling rates and operating at lower fluorescence acquisition rates. 

At the lowest lattice depths, possible limitations to our imaging scheme include the reduced fidelity of RSC due to a lower Lamb-Dicke parameter and off-resonant Raman coupling to higher vibrational bands, an increased susceptibility to photon reabsorption heating \cite{wolf2000} and faster rates of tunneling to neighbouring lattice sites. As we show in Fig. 4, these limitations can be overcome by a suitable choice of fluorescence acquisition rate and Raman cooling rates. Already, the lowest lattice depth ($s \sim 15$) for which we demonstrate nondestructive imaging is more than two orders of magnitude below that required for molasses-based lattice imaging. 

\begin{figure}
\includegraphics[width=2.60in]{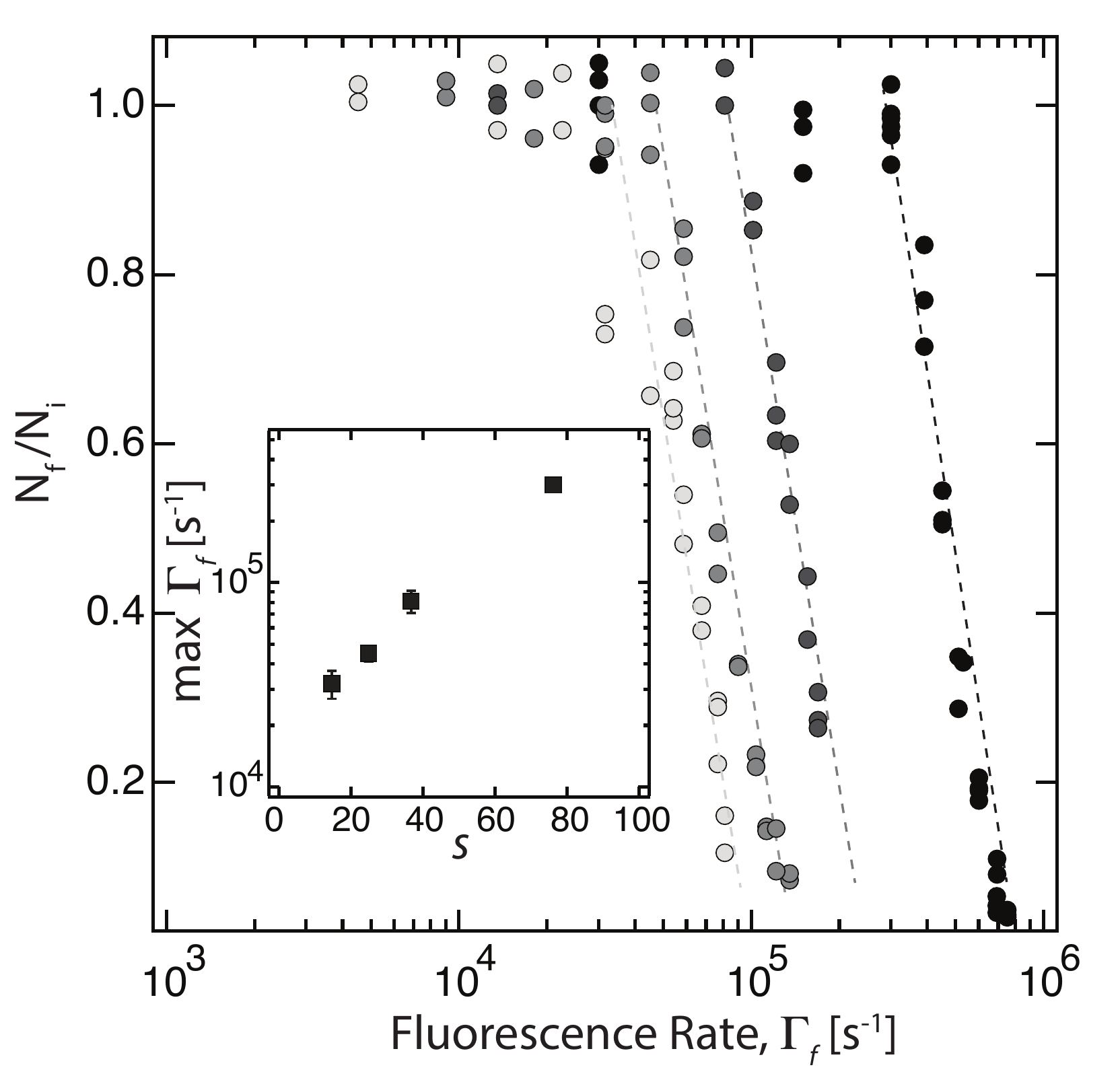}
\caption{The fraction of atoms remaining after the imaging sequence ($N_f/N_i$) {\em vs} fluorescence rates for lattice depths of $U_0/E_r= 14.6, 24.9, 36.7$ and $76.2$ (left to right). 
Inset: An estimate of the maximum fluorescence acquisition rate $\Gamma_{f, max}$ per atom {\em vs} $s = U_0/E_r$.} 
\label{Fig:Fig4}
\end{figure}

In summary, we demonstrate a nondestructive imaging technique for ultracold atoms confined in an optical lattice. The imaging technique is based on extracting fluorescence while simultaneously cooling the atoms to the ground state of the lattice via Raman sideband cooling. 
Using a combination of sideband spectroscopy and time-of-flight imaging, we demonstrate a large operational regime of fluorescence acquisition rates and lattice depths for which the imaging scheme preserves the spatial location, the spin state as well as the vibrational occupation of the atoms. At the largest rates of fluorescence acquisition ($\sim 10^6$ photons/s/atom) and the lowest lattice depths, the main loss mechanism occurs due to tunneling of atoms to occupied lattice sites followed by rapid light-induced loss. By using the light-induced loss as a diagnostic measure of tunneling, we show that this limitation can be alleviated by a suitable choice of fluorescence and Raman cooling rates. That said, we note that the imaging scheme demonstrated here does lead to light-induced loss in lattice sites occupied by multiple atoms. In this regard, it is similar to molasses-based imaging in its sensitivity to the `parity' of lattice occupancy.

Our imaging technique represents a powerful extension of lattice imaging to the nondestructive control and measurement of lattice gases. As such, it is an enabling technique to extend concepts of single particle quantum control and QND measurement \cite{wineland2011,guerlin2007,hume2007} to the context of strongly correlated many-body systems. This scheme also permits the continuous monitoring of the out-of-equilibrium dynamics of ultracold lattice gases. We also note that while used primarily as a diagnostic tool here, spatially resolved coherent two-photon processes such as the setup used for sideband spectroscopy in our work, can also be used for sub-diffraction limited quantum control of the lattice gas \cite{johnson1998,gorshkov2008}. Lastly, our imaging scheme is also extendable to atomic species that are less amenable to molasses based lattice imaging as well as to novel lattice geometries \cite{windpassinger2013} where molasses based imaging can be stymied by local polarization gradients.

This work was supported by the ARO MURI on Non-equilibrium Many-body Dynamics (63834-PH-MUR), the DARPA QuASAR program through a grant from the ARO and the Cornell Center for Materials Research with funding from the NSF MRSEC program (DMR-1120296). L. M. A. acknowledges support from a NSF graduate research fellowship. M. V. acknowledges support from the Alfred P. Sloan Foundation. 

\bibliography{Imagingbib,Imagingnotes}

\begin{thebibliography}{28}%
\makeatletter
\providecommand \@ifxundefined [1]{%
 \@ifx{#1\undefined}
}%
\providecommand \@ifnum [1]{%
 \ifnum #1\expandafter \@firstoftwo
 \else \expandafter \@secondoftwo
 \fi
}%
\providecommand \@ifx [1]{%
 \ifx #1\expandafter \@firstoftwo
 \else \expandafter \@secondoftwo
 \fi
}%
\providecommand \natexlab [1]{#1}%
\providecommand \enquote  [1]{``#1''}%
\providecommand \bibnamefont  [1]{#1}%
\providecommand \bibfnamefont [1]{#1}%
\providecommand \citenamefont [1]{#1}%
\providecommand \href@noop [0]{\@secondoftwo}%
\providecommand \href [0]{\begingroup \@sanitize@url \@href}%
\providecommand \@href[1]{\@@startlink{#1}\@@href}%
\providecommand \@@href[1]{\endgroup#1\@@endlink}%
\providecommand \@sanitize@url [0]{\catcode `\\12\catcode `\$12\catcode
  `\&12\catcode `\#12\catcode `\^12\catcode `\_12\catcode `\%12\relax}%
\providecommand \@@startlink[1]{}%
\providecommand \@@endlink[0]{}%
\providecommand \url  [0]{\begingroup\@sanitize@url \@url }%
\providecommand \@url [1]{\endgroup\@href {#1}{\urlprefix }}%
\providecommand \urlprefix  [0]{URL }%
\providecommand \Eprint [0]{\href }%
\providecommand \doibase [0]{http://dx.doi.org/}%
\providecommand \selectlanguage [0]{\@gobble}%
\providecommand \bibinfo  [0]{\@secondoftwo}%
\providecommand \bibfield  [0]{\@secondoftwo}%
\providecommand \translation [1]{[#1]}%
\providecommand \BibitemOpen [0]{}%
\providecommand \bibitemStop [0]{}%
\providecommand \bibitemNoStop [0]{.\EOS\space}%
\providecommand \EOS [0]{\spacefactor3000\relax}%
\providecommand \BibitemShut  [1]{\csname bibitem#1\endcsname}%
\let\auto@bib@innerbib\@empty
\bibitem [{\citenamefont {Lewenstein}\ \emph {et~al.}(2007)\citenamefont
  {Lewenstein}, \citenamefont {Sanpera}, \citenamefont {Ahufinger},
  \citenamefont {Damski}, \citenamefont {Sen},\ and\ \citenamefont
  {Sen}}]{lewenstein2007}%
  \BibitemOpen
  \bibfield  {author} {\bibinfo {author} {\bibfnamefont {M.}~\bibnamefont
  {Lewenstein}}, \bibinfo {author} {\bibfnamefont {A.}~\bibnamefont {Sanpera}},
  \bibinfo {author} {\bibfnamefont {V.}~\bibnamefont {Ahufinger}}, \bibinfo
  {author} {\bibfnamefont {B.}~\bibnamefont {Damski}}, \bibinfo {author}
  {\bibfnamefont {A.}~\bibnamefont {Sen}}, \ and\ \bibinfo {author}
  {\bibfnamefont {U.}~\bibnamefont {Sen}},\ }\href@noop {} {\bibfield
  {journal} {\bibinfo  {journal} {Adv. Phys.}\ }\textbf {\bibinfo {volume}
  {56}},\ \bibinfo {pages} {243} (\bibinfo {year} {2007})}\BibitemShut
  {NoStop}%
\bibitem [{\citenamefont {Bloch}\ \emph {et~al.}(2008)\citenamefont {Bloch},
  \citenamefont {Dalibard},\ and\ \citenamefont {Zwerger}}]{blochrmp2008}%
  \BibitemOpen
  \bibfield  {author} {\bibinfo {author} {\bibfnamefont {I.}~\bibnamefont
  {Bloch}}, \bibinfo {author} {\bibfnamefont {J.}~\bibnamefont {Dalibard}}, \
  and\ \bibinfo {author} {\bibfnamefont {W.}~\bibnamefont {Zwerger}},\
  }\href@noop {} {\bibfield  {journal} {\bibinfo  {journal} {Rev. Mod. Phys.}\
  }\textbf {\bibinfo {volume} {80}},\ \bibinfo {pages} {885} (\bibinfo {year}
  {2008})}\BibitemShut {NoStop}%
\bibitem [{\citenamefont {Polkovnikov}\ \emph {et~al.}(2011)\citenamefont
  {Polkovnikov}, \citenamefont {Sengupta}, \citenamefont {Silva},\ and\
  \citenamefont {Vengalattore}}]{polkovnikov2011}%
  \BibitemOpen
  \bibfield  {author} {\bibinfo {author} {\bibfnamefont {A.}~\bibnamefont
  {Polkovnikov}}, \bibinfo {author} {\bibfnamefont {K.}~\bibnamefont
  {Sengupta}}, \bibinfo {author} {\bibfnamefont {A.}~\bibnamefont {Silva}}, \
  and\ \bibinfo {author} {\bibfnamefont {M.}~\bibnamefont {Vengalattore}},\
  }\href@noop {} {\bibfield  {journal} {\bibinfo  {journal} {Rev. Mod. Phys.}\
  }\textbf {\bibinfo {volume} {83}},\ \bibinfo {pages} {863} (\bibinfo {year}
  {2011})}\BibitemShut {NoStop}%
\bibitem [{\citenamefont {Bloom}\ \emph {et~al.}(2014)\citenamefont {Bloom},
  \citenamefont {Nicholson}, \citenamefont {Williams}, \citenamefont
  {Campbell}, \citenamefont {Bishof}, \citenamefont {Zhang}, \citenamefont
  {Zhang}, \citenamefont {Bromley},\ and\ \citenamefont {Ye}}]{bloom2014}%
  \BibitemOpen
  \bibfield  {author} {\bibinfo {author} {\bibfnamefont {B.~J.}\ \bibnamefont
  {Bloom}}, \bibinfo {author} {\bibfnamefont {T.~L.}\ \bibnamefont
  {Nicholson}}, \bibinfo {author} {\bibfnamefont {J.~R.}\ \bibnamefont
  {Williams}}, \bibinfo {author} {\bibfnamefont {S.~L.}\ \bibnamefont
  {Campbell}}, \bibinfo {author} {\bibfnamefont {M.}~\bibnamefont {Bishof}},
  \bibinfo {author} {\bibfnamefont {X.}~\bibnamefont {Zhang}}, \bibinfo
  {author} {\bibfnamefont {W.}~\bibnamefont {Zhang}}, \bibinfo {author}
  {\bibfnamefont {S.~L.}\ \bibnamefont {Bromley}}, \ and\ \bibinfo {author}
  {\bibfnamefont {J.}~\bibnamefont {Ye}},\ }\href@noop {} {\bibfield  {journal}
  {\bibinfo  {journal} {Nature}\ }\textbf {\bibinfo {volume} {506}},\ \bibinfo
  {pages} {71} (\bibinfo {year} {2014})}\BibitemShut {NoStop}%
\bibitem [{\citenamefont {Bakr}\ \emph {et~al.}(2009)\citenamefont {Bakr},
  \citenamefont {Gillen}, \citenamefont {Peng}, \citenamefont {F\"{o}lling},\
  and\ \citenamefont {Greiner}}]{bakr2009}%
  \BibitemOpen
  \bibfield  {author} {\bibinfo {author} {\bibfnamefont {W.~S.}\ \bibnamefont
  {Bakr}}, \bibinfo {author} {\bibfnamefont {J.~I.}\ \bibnamefont {Gillen}},
  \bibinfo {author} {\bibfnamefont {A.}~\bibnamefont {Peng}}, \bibinfo {author}
  {\bibfnamefont {S.}~\bibnamefont {F\"{o}lling}}, \ and\ \bibinfo {author}
  {\bibfnamefont {M.}~\bibnamefont {Greiner}},\ }\href@noop {} {\bibfield
  {journal} {\bibinfo  {journal} {Nature}\ }\textbf {\bibinfo {volume} {462}},\
  \bibinfo {pages} {74} (\bibinfo {year} {2009})}\BibitemShut {NoStop}%
\bibitem [{\citenamefont {Gemelke}\ \emph {et~al.}(2009)\citenamefont
  {Gemelke}, \citenamefont {Zhang}, \citenamefont {Hung},\ and\ \citenamefont
  {Chin}}]{gemelke2009}%
  \BibitemOpen
  \bibfield  {author} {\bibinfo {author} {\bibfnamefont {N.}~\bibnamefont
  {Gemelke}}, \bibinfo {author} {\bibfnamefont {X.}~\bibnamefont {Zhang}},
  \bibinfo {author} {\bibfnamefont {C.~L.}\ \bibnamefont {Hung}}, \ and\
  \bibinfo {author} {\bibfnamefont {C.}~\bibnamefont {Chin}},\ }\href@noop {}
  {\bibfield  {journal} {\bibinfo  {journal} {Nature}\ }\textbf {\bibinfo
  {volume} {460}},\ \bibinfo {pages} {995} (\bibinfo {year}
  {2009})}\BibitemShut {NoStop}%
\bibitem [{\citenamefont {W\"{u}rtz}\ \emph {et~al.}(2009)\citenamefont
  {W\"{u}rtz}, \citenamefont {Langen}, \citenamefont {Gericke}, \citenamefont
  {Koglbauer},\ and\ \citenamefont {Ott}}]{wurtz2009}%
  \BibitemOpen
  \bibfield  {author} {\bibinfo {author} {\bibfnamefont {P.}~\bibnamefont
  {W\"{u}rtz}}, \bibinfo {author} {\bibfnamefont {T.}~\bibnamefont {Langen}},
  \bibinfo {author} {\bibfnamefont {T.}~\bibnamefont {Gericke}}, \bibinfo
  {author} {\bibfnamefont {A.}~\bibnamefont {Koglbauer}}, \ and\ \bibinfo
  {author} {\bibfnamefont {H.}~\bibnamefont {Ott}},\ }\href@noop {} {\bibfield
  {journal} {\bibinfo  {journal} {Phys. Rev. Lett.}\ }\textbf {\bibinfo
  {volume} {103}},\ \bibinfo {pages} {080404} (\bibinfo {year}
  {2009})}\BibitemShut {NoStop}%
\bibitem [{\citenamefont {Sherson}\ \emph {et~al.}(2010)\citenamefont
  {Sherson}, \citenamefont {Weitenberg}, \citenamefont {Endres}, \citenamefont
  {Chaneau}, \citenamefont {Bloch},\ and\ \citenamefont {Kuhr}}]{sherson2010}%
  \BibitemOpen
  \bibfield  {author} {\bibinfo {author} {\bibfnamefont {J.~F.}\ \bibnamefont
  {Sherson}}, \bibinfo {author} {\bibfnamefont {C.}~\bibnamefont {Weitenberg}},
  \bibinfo {author} {\bibfnamefont {M.}~\bibnamefont {Endres}}, \bibinfo
  {author} {\bibfnamefont {M.}~\bibnamefont {Chaneau}}, \bibinfo {author}
  {\bibfnamefont {I.}~\bibnamefont {Bloch}}, \ and\ \bibinfo {author}
  {\bibfnamefont {S.}~\bibnamefont {Kuhr}},\ }\href@noop {} {\bibfield
  {journal} {\bibinfo  {journal} {Nature}\ }\textbf {\bibinfo {volume} {467}},\
  \bibinfo {pages} {68} (\bibinfo {year} {2010})}\BibitemShut {NoStop}%
\bibitem [{\citenamefont {Nelson}\ \emph {et~al.}(2007)\citenamefont {Nelson},
  \citenamefont {Li},\ and\ \citenamefont {Weiss}}]{nelson2007}%
  \BibitemOpen
  \bibfield  {author} {\bibinfo {author} {\bibfnamefont {K.~D.}\ \bibnamefont
  {Nelson}}, \bibinfo {author} {\bibfnamefont {X.}~\bibnamefont {Li}}, \ and\
  \bibinfo {author} {\bibfnamefont {D.~S.}\ \bibnamefont {Weiss}},\ }\href@noop
  {} {\bibfield  {journal} {\bibinfo  {journal} {Nat. Phys.}\ }\textbf
  {\bibinfo {volume} {3}},\ \bibinfo {pages} {556} (\bibinfo {year}
  {2007})}\BibitemShut {NoStop}%
\bibitem [{\citenamefont {Vuleti\'{c}}\ \emph {et~al.}(1998)\citenamefont
  {Vuleti\'{c}}, \citenamefont {Chin}, \citenamefont {Kerman},\ and\
  \citenamefont {Chu}}]{vuletic1998}%
  \BibitemOpen
  \bibfield  {author} {\bibinfo {author} {\bibfnamefont {V.}~\bibnamefont
  {Vuleti\'{c}}}, \bibinfo {author} {\bibfnamefont {C.}~\bibnamefont {Chin}},
  \bibinfo {author} {\bibfnamefont {A.~J.}\ \bibnamefont {Kerman}}, \ and\
  \bibinfo {author} {\bibfnamefont {S.}~\bibnamefont {Chu}},\ }\href@noop {}
  {\bibfield  {journal} {\bibinfo  {journal} {Phys. Rev. Lett.}\ }\textbf
  {\bibinfo {volume} {81}},\ \bibinfo {pages} {5768} (\bibinfo {year}
  {1998})}\BibitemShut {NoStop}%
\bibitem [{\citenamefont {Hamann}\ \emph {et~al.}(1998)\citenamefont {Hamann},
  \citenamefont {Haycock}, \citenamefont {Klose}, \citenamefont {Pax},
  \citenamefont {Deutsch},\ and\ \citenamefont {Jessen}}]{hamann1998}%
  \BibitemOpen
  \bibfield  {author} {\bibinfo {author} {\bibfnamefont {S.~E.}\ \bibnamefont
  {Hamann}}, \bibinfo {author} {\bibfnamefont {D.~L.}\ \bibnamefont {Haycock}},
  \bibinfo {author} {\bibfnamefont {G.}~\bibnamefont {Klose}}, \bibinfo
  {author} {\bibfnamefont {P.~H.}\ \bibnamefont {Pax}}, \bibinfo {author}
  {\bibfnamefont {I.~H.}\ \bibnamefont {Deutsch}}, \ and\ \bibinfo {author}
  {\bibfnamefont {P.~S.}\ \bibnamefont {Jessen}},\ }\href@noop {} {\bibfield
  {journal} {\bibinfo  {journal} {Phys. Rev. Lett.}\ }\textbf {\bibinfo
  {volume} {80}},\ \bibinfo {pages} {4149} (\bibinfo {year}
  {1998})}\BibitemShut {NoStop}%
\bibitem [{\citenamefont {Han}\ \emph {et~al.}(2000)\citenamefont {Han},
  \citenamefont {Wolf}, \citenamefont {Oliver}, \citenamefont {McCormick},
  \citenamefont {DePue},\ and\ \citenamefont {Weiss}}]{han2000}%
  \BibitemOpen
  \bibfield  {author} {\bibinfo {author} {\bibfnamefont {D.~J.}\ \bibnamefont
  {Han}}, \bibinfo {author} {\bibfnamefont {S.}~\bibnamefont {Wolf}}, \bibinfo
  {author} {\bibfnamefont {S.}~\bibnamefont {Oliver}}, \bibinfo {author}
  {\bibfnamefont {C.}~\bibnamefont {McCormick}}, \bibinfo {author}
  {\bibfnamefont {M.~T.}\ \bibnamefont {DePue}}, \ and\ \bibinfo {author}
  {\bibfnamefont {D.~S.}\ \bibnamefont {Weiss}},\ }\href@noop {} {\bibfield
  {journal} {\bibinfo  {journal} {Phys. Rev. Lett.}\ }\textbf {\bibinfo
  {volume} {85}},\ \bibinfo {pages} {724} (\bibinfo {year} {2000})}\BibitemShut
  {NoStop}%
\bibitem [{\citenamefont {Kerman}\ \emph {et~al.}(2000)\citenamefont {Kerman},
  \citenamefont {Vuleti\'{c}}, \citenamefont {Chin},\ and\ \citenamefont
  {Chu}}]{kerman2000}%
  \BibitemOpen
  \bibfield  {author} {\bibinfo {author} {\bibfnamefont {A.~J.}\ \bibnamefont
  {Kerman}}, \bibinfo {author} {\bibfnamefont {V.}~\bibnamefont {Vuleti\'{c}}},
  \bibinfo {author} {\bibfnamefont {C.}~\bibnamefont {Chin}}, \ and\ \bibinfo
  {author} {\bibfnamefont {S.}~\bibnamefont {Chu}},\ }\href@noop {} {\bibfield
  {journal} {\bibinfo  {journal} {Phys. Rev. Lett.}\ }\textbf {\bibinfo
  {volume} {84}},\ \bibinfo {pages} {439} (\bibinfo {year} {2000})}\BibitemShut
  {NoStop}%
\bibitem [{\citenamefont {Monroe}\ \emph {et~al.}(1995)\citenamefont {Monroe},
  \citenamefont {Meekhof}, \citenamefont {King}, \citenamefont {Jefferts},
  \citenamefont {Itano}, \citenamefont {Wineland},\ and\ \citenamefont
  {Gould}}]{monroe1995}%
  \BibitemOpen
  \bibfield  {author} {\bibinfo {author} {\bibfnamefont {C.}~\bibnamefont
  {Monroe}}, \bibinfo {author} {\bibfnamefont {D.~M.}\ \bibnamefont {Meekhof}},
  \bibinfo {author} {\bibfnamefont {B.~E.}\ \bibnamefont {King}}, \bibinfo
  {author} {\bibfnamefont {S.~R.}\ \bibnamefont {Jefferts}}, \bibinfo {author}
  {\bibfnamefont {W.~M.}\ \bibnamefont {Itano}}, \bibinfo {author}
  {\bibfnamefont {D.~J.}\ \bibnamefont {Wineland}}, \ and\ \bibinfo {author}
  {\bibfnamefont {P.}~\bibnamefont {Gould}},\ }\href@noop {} {\bibfield
  {journal} {\bibinfo  {journal} {Phys. Rev. Lett.}\ }\textbf {\bibinfo
  {volume} {75}},\ \bibinfo {pages} {4011} (\bibinfo {year}
  {1995})}\BibitemShut {NoStop}%
\bibitem [{tof()}]{toffootnote}%
  \BibitemOpen
  \href@noop {} {}\bibinfo {note} {For very large atom numbers ($\sim 10^8$)
  where the atomic cloud size is commensurate with that of the lattice waist,
  the time-of-flight measurements indicate slightly elevated temperatures in
  comparison to that extracted from the local sideband spectra. This is
  presumably due to the inhomogeneous variation of lattice frequencies across
  the atomic distribution and a reduced efficacy of RSC for atoms on the
  periphery. In such cases, we rely on the sideband spectra for a more reliable
  estimate of the temperature.}\BibitemShut {Stop}%
\bibitem [{\citenamefont {Cohen-Tannoudji}\ \emph {et~al.}(1998)\citenamefont
  {Cohen-Tannoudji}, \citenamefont {Dupont-Roc},\ and\ \citenamefont
  {Grynberg}}]{cct1998}%
  \BibitemOpen
  \bibfield  {author} {\bibinfo {author} {\bibfnamefont {C.}~\bibnamefont
  {Cohen-Tannoudji}}, \bibinfo {author} {\bibfnamefont {J.}~\bibnamefont
  {Dupont-Roc}}, \ and\ \bibinfo {author} {\bibfnamefont {G.}~\bibnamefont
  {Grynberg}},\ }\href@noop {} {\emph {\bibinfo {title} {Atom Photon
  Interactions}}}\ (\bibinfo  {publisher} {Wiley-Interscience, New York},\
  \bibinfo {year} {1998})\BibitemShut {NoStop}%
\bibitem [{rat()}]{ratenote}%
  \BibitemOpen
  \href@noop {} {}\bibinfo {note} {The fluorescence rate is calibrated by the
  measured rate of Raman fluorescence acquired by the camera, and a geometrical
  estimate of the numerical aperture of our imaging system. This calibration is
  very close to the rate estimated from the measured intensity and detuning of
  the fluorescence beam.}\BibitemShut {Stop}%
\bibitem [{\citenamefont {Syassen}\ \emph {et~al.}(2008)\citenamefont
  {Syassen}, \citenamefont {Bauer}, \citenamefont {Lettner}, \citenamefont
  {Volz}, \citenamefont {Dietze}, \citenamefont {Garcia-Ripoll}, \citenamefont
  {Cirac}, \citenamefont {Rempe},\ and\ \citenamefont
  {D\"{u}rr}}]{syassen2008}%
  \BibitemOpen
  \bibfield  {author} {\bibinfo {author} {\bibfnamefont {N.}~\bibnamefont
  {Syassen}}, \bibinfo {author} {\bibfnamefont {D.~M.}\ \bibnamefont {Bauer}},
  \bibinfo {author} {\bibfnamefont {M.}~\bibnamefont {Lettner}}, \bibinfo
  {author} {\bibfnamefont {T.}~\bibnamefont {Volz}}, \bibinfo {author}
  {\bibfnamefont {D.}~\bibnamefont {Dietze}}, \bibinfo {author} {\bibfnamefont
  {J.~J.}\ \bibnamefont {Garcia-Ripoll}}, \bibinfo {author} {\bibfnamefont
  {J.~I.}\ \bibnamefont {Cirac}}, \bibinfo {author} {\bibfnamefont
  {G.}~\bibnamefont {Rempe}}, \ and\ \bibinfo {author} {\bibfnamefont
  {S.}~\bibnamefont {D\"{u}rr}},\ }\href@noop {} {\bibfield  {journal}
  {\bibinfo  {journal} {Science}\ }\textbf {\bibinfo {volume} {320}},\ \bibinfo
  {pages} {1329} (\bibinfo {year} {2008})}\BibitemShut {NoStop}%
\bibitem [{\citenamefont {Yan}\ \emph {et~al.}(2013)\citenamefont {Yan},
  \citenamefont {Moses}, \citenamefont {Gadway}, \citenamefont {Covey},
  \citenamefont {Hazzard}, \citenamefont {Rey}, \citenamefont {Jin},\ and\
  \citenamefont {Ye}}]{yan2013}%
  \BibitemOpen
  \bibfield  {author} {\bibinfo {author} {\bibfnamefont {B.}~\bibnamefont
  {Yan}}, \bibinfo {author} {\bibfnamefont {S.~A.}\ \bibnamefont {Moses}},
  \bibinfo {author} {\bibfnamefont {B.}~\bibnamefont {Gadway}}, \bibinfo
  {author} {\bibfnamefont {J.~P.}\ \bibnamefont {Covey}}, \bibinfo {author}
  {\bibfnamefont {K.~R.~A.}\ \bibnamefont {Hazzard}}, \bibinfo {author}
  {\bibfnamefont {A.~M.}\ \bibnamefont {Rey}}, \bibinfo {author} {\bibfnamefont
  {D.~S.}\ \bibnamefont {Jin}}, \ and\ \bibinfo {author} {\bibfnamefont
  {J.}~\bibnamefont {Ye}},\ }\href@noop {} {\bibfield  {journal} {\bibinfo
  {journal} {Nature}\ }\textbf {\bibinfo {volume} {501}},\ \bibinfo {pages}
  {521} (\bibinfo {year} {2013})}\BibitemShut {NoStop}%
\bibitem [{\citenamefont {Jaksch}\ \emph {et~al.}(1998)\citenamefont {Jaksch},
  \citenamefont {Bruder}, \citenamefont {Cirac}, \citenamefont {Gardiner},\
  and\ \citenamefont {Zoller}}]{jaksch1998}%
  \BibitemOpen
  \bibfield  {author} {\bibinfo {author} {\bibfnamefont {D.}~\bibnamefont
  {Jaksch}}, \bibinfo {author} {\bibfnamefont {C.}~\bibnamefont {Bruder}},
  \bibinfo {author} {\bibfnamefont {J.~I.}\ \bibnamefont {Cirac}}, \bibinfo
  {author} {\bibfnamefont {C.~W.}\ \bibnamefont {Gardiner}}, \ and\ \bibinfo
  {author} {\bibfnamefont {P.}~\bibnamefont {Zoller}},\ }\href@noop {}
  {\bibfield  {journal} {\bibinfo  {journal} {Phys. Rev. Lett.}\ }\textbf
  {\bibinfo {volume} {81}},\ \bibinfo {pages} {3108} (\bibinfo {year}
  {1998})}\BibitemShut {NoStop}%
\bibitem [{\citenamefont {Zwerger}(2003)}]{zwerger2003}%
  \BibitemOpen
  \bibfield  {author} {\bibinfo {author} {\bibfnamefont {W.}~\bibnamefont
  {Zwerger}},\ }\href@noop {} {\bibfield  {journal} {\bibinfo  {journal} {J.
  Opt. B}\ }\textbf {\bibinfo {volume} {5}},\ \bibinfo {pages} {S9} (\bibinfo
  {year} {2003})}\BibitemShut {NoStop}%
\bibitem [{\citenamefont {Wolf}\ \emph {et~al.}(2000)\citenamefont {Wolf},
  \citenamefont {Oliver},\ and\ \citenamefont {Weiss}}]{wolf2000}%
  \BibitemOpen
  \bibfield  {author} {\bibinfo {author} {\bibfnamefont {S.}~\bibnamefont
  {Wolf}}, \bibinfo {author} {\bibfnamefont {S.~J.}\ \bibnamefont {Oliver}}, \
  and\ \bibinfo {author} {\bibfnamefont {D.~S.}\ \bibnamefont {Weiss}},\
  }\href@noop {} {\bibfield  {journal} {\bibinfo  {journal} {Phys. Rev. Lett.}\
  }\textbf {\bibinfo {volume} {85}},\ \bibinfo {pages} {4249} (\bibinfo {year}
  {2000})}\BibitemShut {NoStop}%
\bibitem [{\citenamefont {Wineland}\ and\ \citenamefont
  {Liebfried}(2011)}]{wineland2011}%
  \BibitemOpen
  \bibfield  {author} {\bibinfo {author} {\bibfnamefont {D.~J.}\ \bibnamefont
  {Wineland}}\ and\ \bibinfo {author} {\bibfnamefont {D.}~\bibnamefont
  {Liebfried}},\ }\href@noop {} {\bibfield  {journal} {\bibinfo  {journal}
  {Laser Phys. Lett.}\ }\textbf {\bibinfo {volume} {8}},\ \bibinfo {pages}
  {175} (\bibinfo {year} {2011})}\BibitemShut {NoStop}%
\bibitem [{\citenamefont {Guerlin}\ \emph {et~al.}(2007)\citenamefont
  {Guerlin}, \citenamefont {Bernu}, \citenamefont {Del\'{e}glise},
  \citenamefont {Sayrin}, \citenamefont {Gleyzes}, \citenamefont {Kuhr},
  \citenamefont {Brune}, \citenamefont {Raimond},\ and\ \citenamefont
  {Haroche}}]{guerlin2007}%
  \BibitemOpen
  \bibfield  {author} {\bibinfo {author} {\bibfnamefont {C.}~\bibnamefont
  {Guerlin}}, \bibinfo {author} {\bibfnamefont {J.}~\bibnamefont {Bernu}},
  \bibinfo {author} {\bibfnamefont {S.}~\bibnamefont {Del\'{e}glise}}, \bibinfo
  {author} {\bibfnamefont {C.}~\bibnamefont {Sayrin}}, \bibinfo {author}
  {\bibfnamefont {S.}~\bibnamefont {Gleyzes}}, \bibinfo {author} {\bibfnamefont
  {S.}~\bibnamefont {Kuhr}}, \bibinfo {author} {\bibfnamefont {M.}~\bibnamefont
  {Brune}}, \bibinfo {author} {\bibfnamefont {J.~M.}\ \bibnamefont {Raimond}},
  \ and\ \bibinfo {author} {\bibfnamefont {S.}~\bibnamefont {Haroche}},\
  }\href@noop {} {\bibfield  {journal} {\bibinfo  {journal} {Nature}\ }\textbf
  {\bibinfo {volume} {448}},\ \bibinfo {pages} {889} (\bibinfo {year}
  {2007})}\BibitemShut {NoStop}%
\bibitem [{\citenamefont {Hume}\ \emph {et~al.}(2007)\citenamefont {Hume},
  \citenamefont {Rosenband},\ and\ \citenamefont {Wineland}}]{hume2007}%
  \BibitemOpen
  \bibfield  {author} {\bibinfo {author} {\bibfnamefont {D.~B.}\ \bibnamefont
  {Hume}}, \bibinfo {author} {\bibfnamefont {T.}~\bibnamefont {Rosenband}}, \
  and\ \bibinfo {author} {\bibfnamefont {D.~J.}\ \bibnamefont {Wineland}},\
  }\href@noop {} {\bibfield  {journal} {\bibinfo  {journal} {Phys. Rev. Lett.}\
  }\textbf {\bibinfo {volume} {99}},\ \bibinfo {pages} {120502} (\bibinfo
  {year} {2007})}\BibitemShut {NoStop}%
\bibitem [{\citenamefont {Johnson}\ \emph {et~al.}(1998)\citenamefont
  {Johnson}, \citenamefont {Thywissen}, \citenamefont {Dekker}, \citenamefont
  {Berggren}, \citenamefont {Chu}, \citenamefont {Younkin},\ and\ \citenamefont
  {Prentiss}}]{johnson1998}%
  \BibitemOpen
  \bibfield  {author} {\bibinfo {author} {\bibfnamefont {K.~S.}\ \bibnamefont
  {Johnson}}, \bibinfo {author} {\bibfnamefont {J.~H.}\ \bibnamefont
  {Thywissen}}, \bibinfo {author} {\bibfnamefont {N.~H.}\ \bibnamefont
  {Dekker}}, \bibinfo {author} {\bibfnamefont {K.~K.}\ \bibnamefont
  {Berggren}}, \bibinfo {author} {\bibfnamefont {A.~P.}\ \bibnamefont {Chu}},
  \bibinfo {author} {\bibfnamefont {R.}~\bibnamefont {Younkin}}, \ and\
  \bibinfo {author} {\bibfnamefont {M.}~\bibnamefont {Prentiss}},\ }\href@noop
  {} {\bibfield  {journal} {\bibinfo  {journal} {Science}\ }\textbf {\bibinfo
  {volume} {280}},\ \bibinfo {pages} {1583} (\bibinfo {year}
  {1998})}\BibitemShut {NoStop}%
\bibitem [{\citenamefont {Gorshkov}\ \emph {et~al.}(2008)\citenamefont
  {Gorshkov}, \citenamefont {Jiang}, \citenamefont {Greiner}, \citenamefont
  {Zoller},\ and\ \citenamefont {Lukin}}]{gorshkov2008}%
  \BibitemOpen
  \bibfield  {author} {\bibinfo {author} {\bibfnamefont {A.~V.}\ \bibnamefont
  {Gorshkov}}, \bibinfo {author} {\bibfnamefont {L.}~\bibnamefont {Jiang}},
  \bibinfo {author} {\bibfnamefont {M.}~\bibnamefont {Greiner}}, \bibinfo
  {author} {\bibfnamefont {P.}~\bibnamefont {Zoller}}, \ and\ \bibinfo {author}
  {\bibfnamefont {M.~D.}\ \bibnamefont {Lukin}},\ }\href@noop {} {\bibfield
  {journal} {\bibinfo  {journal} {Phys. Rev. Lett.}\ }\textbf {\bibinfo
  {volume} {100}},\ \bibinfo {pages} {093005} (\bibinfo {year}
  {2008})}\BibitemShut {NoStop}%
\bibitem [{\citenamefont {Windpassinger}\ and\ \citenamefont
  {Sengstock}(2013)}]{windpassinger2013}%
  \BibitemOpen
  \bibfield  {author} {\bibinfo {author} {\bibfnamefont {P.}~\bibnamefont
  {Windpassinger}}\ and\ \bibinfo {author} {\bibfnamefont {K.}~\bibnamefont
  {Sengstock}},\ }\href@noop {} {\bibfield  {journal} {\bibinfo  {journal}
  {Rep. Prog. Phys.}\ }\textbf {\bibinfo {volume} {76}},\ \bibinfo {pages}
  {086401} (\bibinfo {year} {2013})}\BibitemShut {NoStop}%
\end{thebibliography}%

\end{document}